\documentclass[11pt]{article}
\pdfoutput=1
\PassOptionsToPackage{table,dvipsnames}{xcolor}  
\usepackage{xcolor}
\usepackage[preprint]{acl}
\usepackage{makecell} 
\usepackage{amsmath}
\usepackage{times}
\usepackage{latexsym}
\usepackage{dsfont}
\usepackage{multirow}
\usepackage[most]{tcolorbox}
\tcbuselibrary{breakable}
\usepackage{euscript}
\usepackage[T1]{fontenc}
\usepackage{amsfonts}
\usepackage{tabularx}
\usepackage{booktabs}
\usepackage{amssymb}
\usepackage{mathrsfs}

\usepackage[utf8]{inputenc}
\usepackage{microtype}

\usepackage{inconsolata}

\usepackage{graphicx}

\title{Logical Consistency is Vital: Neural-Symbolic Information Retrieval\\ for Negative-Constraint Queries}

\author{
  Ganlin Xu$^1$\thanks{Equal contribution.} \quad Zhoujia Zhang$^1$\footnotemark[1] \quad Wangyi Mei$^1$  \quad Jiaqing Liang$^1$\thanks{~Co-corresponding authors.} \quad Weijia Lu$^2$ \\ \textbf{Xiaodong Zhang}$^2$ \quad \textbf{Zhifei Yang}$^2$ \quad \textbf{Xiaofeng Ma}$^2$  \quad \textbf{Yanghua Xiao}$^3$\quad \textbf{Deqing Yang}$^1$\footnotemark[2]\\
  $^1$School of Data Science, Fudan University, Shanghai, China \\ 
  $^2$United Automotive Electronic Systems, Shanghai, China  \\ 
  $^3$College of Computer Science and Artificial Intelligence, Fudan University, Shanghai, China \\
  \{glxu24@m., 24210980075@m., 24210980101@m., liangjiaqing@, shawyh@, yangdeqing@\}fudan.edu.cn\\
  \{alfredwjlu, xiaodong.zhang.chn, maxf0124\}@gmail.com,  420572897@qq.com
}

\begin{document}
\pdfoutput=1
	\maketitle
	\begin{abstract}
		Information retrieval plays a crucial role in resource localization. Current dense retrievers retrieve the relevant documents within a corpus via embedding similarities, which compute similarities between dense vectors mainly depending on word co-occurrence between queries and documents, but overlook the real query intents.
        Thus, they often retrieve numerous irrelevant documents. Particularly in the scenarios of complex queries such as \emph{negative-constraint queries}, their retrieval performance could be catastrophic. To address the issue, we propose a neuro-symbolic information retrieval method, namely \textbf{NS-IR}, that leverages first-order logic (FOL) to optimize the embeddings of naive natural language by considering the \emph{logical consistency} between queries and documents. Specifically, we introduce two novel techniques, \emph{logic alignment} and \emph{connective constraint}, to rerank candidate documents, thereby enhancing retrieval relevance. 
		Furthermore, we construct a new dataset \textbf{NegConstraint} including negative-constraint queries to evaluate our NS-IR's performance on such complex IR scenarios.
		Our extensive experiments demonstrate that NS-IR not only achieves superior zero-shot retrieval performance on web search and low-resource retrieval tasks, but also performs better on negative-constraint queries. Our scource code and dataset are available at \url{https://github.com/xgl-git/NS-IR-main}.
	\end{abstract}
	
	\section{Introduction}
	Information retrieval (IR) tasks aim at obtaining relevant information from large-scale data collection, such as documents and databases. Dense retrieval \cite{karpukhin-etal-2020-dense} is an advanced information retrieval technique focusing on semantic embedding similarities between texts. It has been widely adopted in many applications such as search engines \cite{li2022unsupervised}, question answering \cite{zhao2021distantly} and retrieval-augmented generation (RAG) systems \cite{huly2024old}, offering significant improvements in IR.
	\begin{figure}[t]
		\centering
		\includegraphics[width=1.0\linewidth]{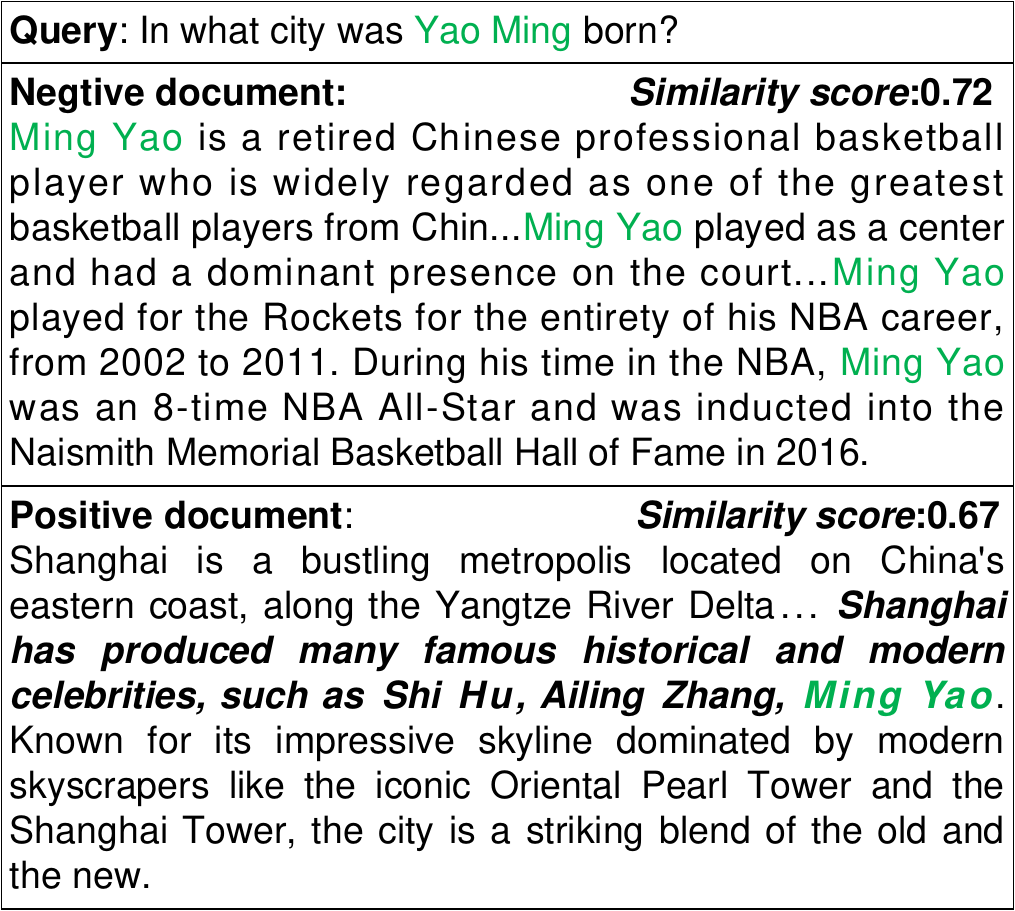}
		\caption{An illustration of BGE-based retrieval. The word marked in green is the co-occurrence word between the query and documents.}\label{fig1}
    \vspace{-0.1cm}
	\end{figure}\\
	\indent The embeddings (representations) generated by dense retrievers (such as BGE \cite{xiao2024cpackpackedresourcesgeneral}) focus on overall semantic similarity, which is capable of handling semantically similar words compared to sparse retrieval (such as BM25 \cite{robertson2009probabilistic}) that uses keyword matching. However, dense retrieval still relies on superficial word co-occurrence between queries and documents. As illustrated in Figure~\ref{fig1}, 
    the negative document has a bigger score than the positive document just because the query's keyword ``Ming Yao'' occurs in the former more frequently.
	Thus, the approach fails to understand the real query intent, thereby retrieving irrelevant documents \cite{wu2024easilyirrelevantinputsskew, fang-etal-2024-enhancing}.\\
	\begin{figure}[h]
		\centering
		\includegraphics[width=\linewidth]{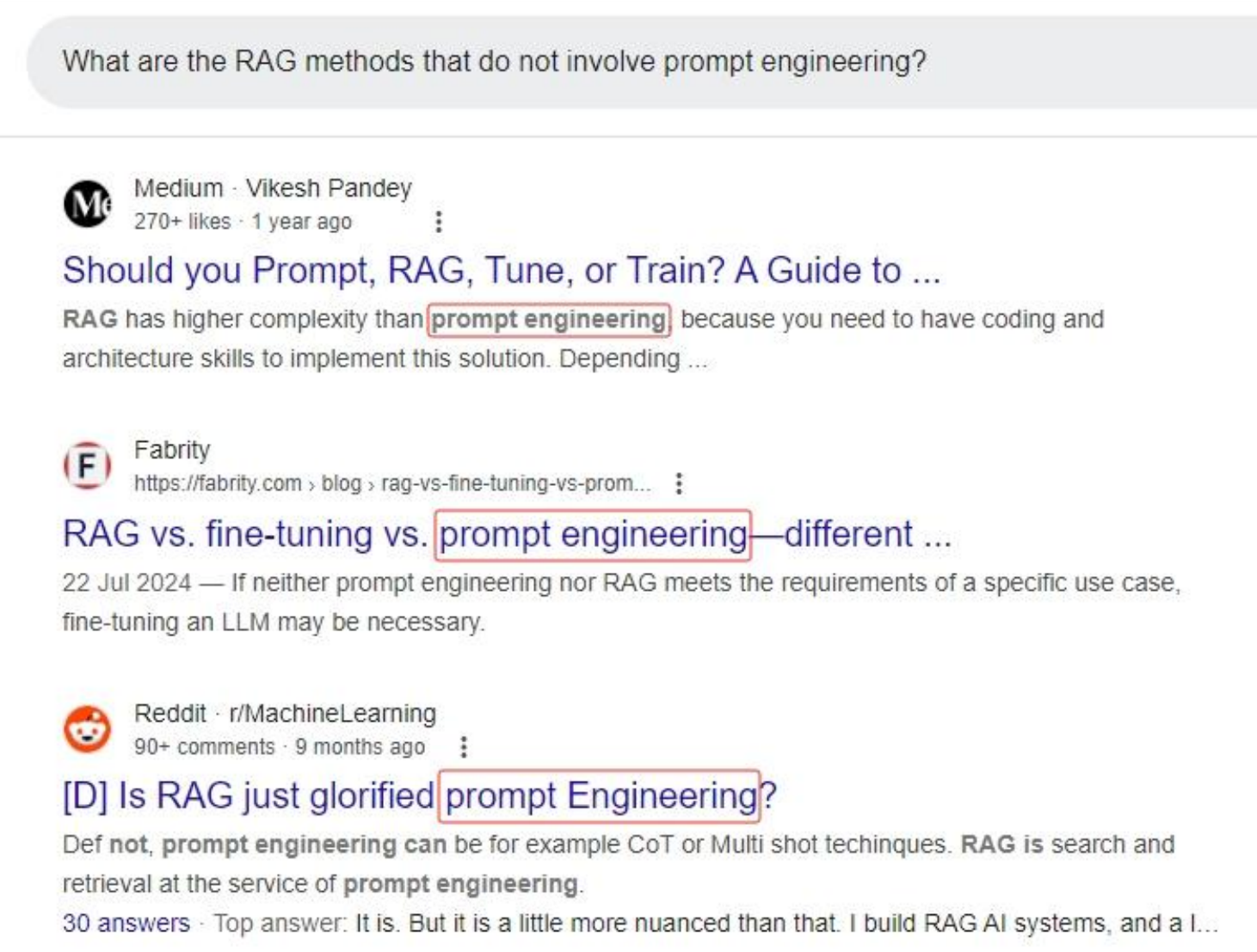}
    \vspace{-0.2cm}
		\caption{A retrieval example of Google search engine.}\label{fig2}
	\end{figure}
	\indent Notably, the retrieval approaches based on word co-occurrence have to face some challenges on complex queries, particularly involving \emph{negative-constraint queries}, due to the neglect of logical consistency. 
	As shown in Figure~\ref{fig2}\footnote{In the example, the search engine's retrieval is based on BM25 algorithm, but dense retrieval produces similar results in negative-constraint queries.}, for a general query ``What are the RAG methods that do not involve prompt engineering?'', the documents returned by a keyword-based search engine often contain the excluded keyword ``prompt engineering'', which are not logically consistent with the query intent\footnote{Although some search techniques in search engines can use `-' to filter keywords (such as ``What are the RAG methods -prompt engineering''), they are not familiar to ordinary users.}. 
    Therefore, understanding real query intents requires ensuring logical consistency between queries and documents besides semantic similarity.
    First-order logic (FOL), as a formal logical system, offers clear logical semantics and expresses complex relations in natural language \cite{barwise1977introduction}. For instance, the FOL of the aforementioned query is ``RAGMethod(x) $\land$ $\neg$InvolvesPromptEngineering(x)'', which clearly expresses the negative semantics through the logical connective `$\neg$'.
	
	In light of this, through investigating the potential impact of FOL on complex logical queries, we propose a \textbf{N}eural-\textbf{S}ymbolic \textbf{I}nformation \textbf{R}etrieval method (NS-IR) to rerank the candidate documents returned by dense retrievers based on FOL, for more accurate retrieval results.
	Specifically, NS-IR first retrieves a set of ordered candidate documents using a dense retriever, then employs large language models (such as GPT-4o) to convert both the query and documents into FOL. To incorporate logical consistency and optimize embeddings of naive natural language (NL), we propose two independent techniques: 1) \emph{Logic alignment}: To incorporate overall logic semantics in FOL into NL representations, we measure distribution differences between NL and FOL embeddings using optimal transport \cite{redko2019optimal} and update the embeddings of queries and documents respectively. 2) \emph{Connective constraint}: To reflect the role of partial words in FOL on logical consistency, we leverage different words in FOL (especially connectives) to render different attentions to words in NL, generating better embeddings with logical semantics. We leverage these two techniques to recalculate similarity scores and rerank candidate documents.
	
	To evaluate the performance of NS-IR in negative-constraint query settings, we have constructed and released a dataset, namely \textbf{NegConstraint}, which contains three types of negative-constraint queries and was sourced from the Wikipedia dump \cite{karpukhin-etal-2020-dense}. The experiments show our NS-IR's superiority over some state-of-the-art (SOTA) baselines on {NegConstraint}, and its potential for handling complex logical queries.\\
	\indent The main contributions of this paper include:
	
	1. To address typical complex logical queries in IR, i.e., \emph{negative-constraint queries}, we propose NS-IR which combines the strengths of NL and FOL to synthesize semantic similarity and logical consistency.
	
	2. We introduce two key techniques: logic alignment (Sec.~\ref{sec4.2}) and connective constraint (Sec.~\ref{sec4.3}), to optimize naive NL embeddings by FOL, and rerank candidate documents to improve retrieval performance.
	
	3. We further release a new dataset {NegConstraint} (Sec.~\ref{sec5}), which can be utilized as a benchmark for negative-constraint queries. Our extensive experimental results on public datasets and {NegConstraint} show that our NS-IR significantly outperforms the SOTA methods on vanilla and negative-constraint queries in zero-shot settings, respectively. Our work in this paper paves the way for future research on complex queries in IR.
	
	\section{Related Work}
	\subsection{Dense Retrieval}
	Dense retrieval has gained significant attention in information retrieval due to its advantages over traditional sparse vector space models. Sparse models represent documents and queries as high-dimensional vectors with mostly zero values \cite{yang2017anserini, chen-etal-2017-reading}. Dense retrieval models encode queries and documents into dense and low-dimensional vectors \cite{karpukhin-etal-2020-dense, cai2022hyper}, which capture semantic similarity instead of match of terms, thus significantly outperforming sparse approaches. Relevant studies mainly focus on improving training approach \cite{qu2020rocketqa}, distillation \cite{zhang2023led} and pre-training \cite{shen2022lexmae} for retrieval.\\ 
	\indent Many studies adopt a transfer learning framework where dense retrieval models are trained on high-resource passage retrieval datasets such as MS-MARCO \cite{bajaj2018human} and then evaluated on queries from new tasks. However, collecting such large-scale corpora is both time-consuming and labor-intensive. Recent work has introduced zero-shot dense retrieval settings, which eliminates the need for relevance labels between queries and documents \cite{gao2022precise}. Our work follows the zero-shot unsupervised setup for all experiments. 
	\subsection{Optimal Transport in NLP}\label{optimaltransport}
	Optimal transport (OT) has been employed in various NLP tasks, where alignment exists implicitly or explicitly. The typical applications of OT include evaluating the similarity between sentences and documents \cite{wang2022unsupervised,mysore2021multi,lee2022toward} or aligning cross-domain representations across different modalities \cite{zhou-etal-2023-cmot, qiu-etal-2023-sccs}. The evaluation mechanism can be integrated as a penalty term into language generation models \cite{chen2019improving, zhang2020topic, li-etal-2020-improving-text}. Moreover, OT effectively handles imbalanced word alignment, including both explicit alignment and null alignment \cite{arase-etal-2023-unbalanced}. Inspired by unsupervised word alignment \cite{arase-etal-2023-unbalanced, huang-etal-2024-ottawa}, we utilize the alignment matrix to measure distribution differences between natural language and first-order logic, enabling natural language to better focus on the logical semantics inherent in first-order logic. 
	\subsection{NL-FOL Translation}
	NL-FOL (Natural Language to First-Order Logic) translation has long been a challenge in both natural language processing (NLP) and formal logic research. Traditionally, NL-FOL translation has been approached through rule-based methods \cite{abzianidze2017langpro}. However, due to the inherent complexity of natural language, these methods struggle to scale to real-world applications. As a result, traditional logic-based reasoning techniques have lost popularity due to limited scalability and coverage.\\ 
	\indent The recent breakthroughs in LLMs have reignited interest in logic, bringing it back to the forefront of reasoning tasks. One promising strategy to leverage the power of LLMs is to translate NL statements, such as premises and conclusions in textual entailment tasks, into first-order logic (FOL) formulas via in-context learning. These symbolic representations can be passed to Symbolic Mathematical Theory (SMT) solvers \cite{olausson-etal-2023-linc, xu-etal-2024-faithful} or used to make veracity predictions and generate explanations \cite{wang-shu-2023-explainable}. In the context, \cite{yang-etal-2024-harnessing} presents a NL-FOL dataset MALLs of 28K diverse and verified sentence-level pairs collected from GPT4, and a translator LOGICLLAMA, a LLaMA2-7B/13B fine-tuned on MALLS for NL-FOL translation.  In this paper, we use LLMs as translators to implement NL-FOL translation. 
	
	\begin{figure*}[t]
		\centering	\includegraphics[width=1.0\linewidth]{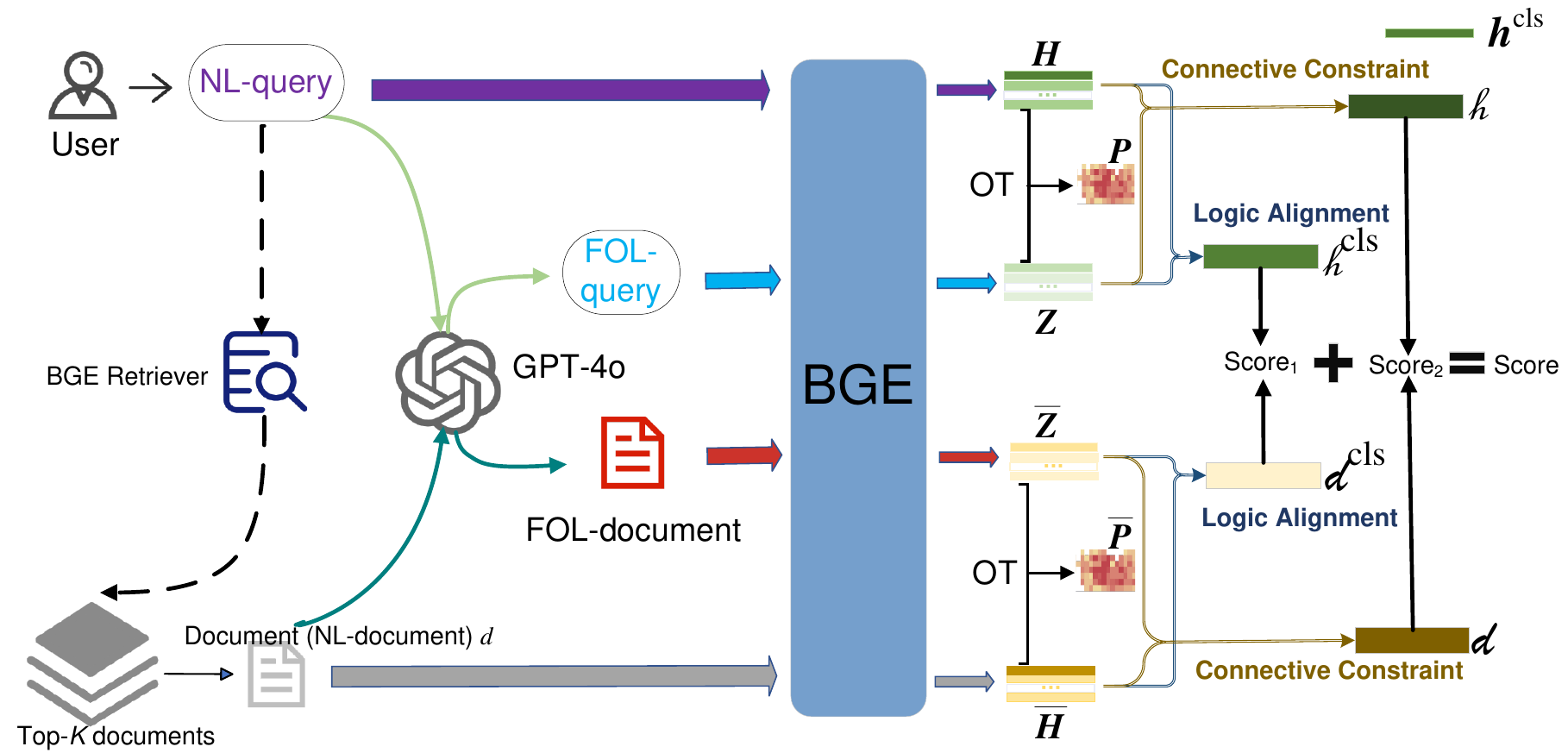}
		\caption{The pipeline of our proposed NS-IR. Dashed arrows represent the retrieval stage. In the figure, only one document \emph{d} in top-\emph{K} documents is encoded, but actually, all top-\emph{K} documents are encoded together.}\label{fig3}
	\end{figure*}
	
	\section{Preliminaries}
	\subsection{Task Formulation}
    In this paper, we focus on the task of zero-shot document retrieval, of which the model captures the similarity between queries and documents without model training.
	Given a query \emph{q} and the document set \emph{D} containing multiple documents, the goal of retrievers is to retrieve document \emph{d} that satisfies the user's real search intent. Dense retrieval uses encoders to map \emph{q} and \emph{d} into a pair of dense vectors, whose inner product is leveraged as a similarity function:
	\begin{equation}
	\text{sim}(q, d) = \langle E_q(q), E_d(d) \rangle. \label{eq1}
	\end{equation}
	\indent In this paper, we use the BGE model as query encoder $E_{q}$ and document encoder $E_{d}$, and the embeddings of {CLS} token (the first token of a sequence) denotes dense vectors $E_q(q)$ and $E_d(d)$. Besides, we obtain word embeddings of NL and FOL sequences via BGE, respectively. Let 
    $\textbf{\emph{H}} = [\textbf{\emph{h}}_1, \ldots, \textbf{\emph{h}}_m] (\textbf{\emph{h}}_i \in \mathbb{R}^{d}, 1\leq i\leq m)$ 
    and 
    $\textbf{\emph{Z}} = [\textbf{\emph{z}}_1, \ldots, \textbf{\emph{z}}_n] (\textbf{\emph{z}}_j \in \mathbb{R}^{d}, 1\leq i\leq n)$ 
    be the embeddings of NL-queries and FOL-queries, respectively. We obtain the embeddings of NL-documents and FOL-documents in the same way\footnote{In the paper, NL-query and FOL-query refer to queries in natural language and first-order logic, respectively. Similarly, NL-documents and FOL-documents correspond to documents in natural language and first-order logic, respectively.}. We provide some examples of FOL in Appendix~\ref{folexamples}.
	
	\subsection{Optimal Transport}
	Optimal transport (OT) seeks to find the most efficient way to transport one probability distribution $\mu$  (the source) to another $\nu$ (the target) while minimizing a predefined cost function \cite{redko2019optimal}. Formally, let $\mu$ and  $\nu$ be probability measures on spaces $\mathcal{X}$ and $\mathcal{Y}$, respectively, and let function $c(\cdot, \cdot)$ represent the cost of transporting a unit of mass from points. The following explanation assumes that the source and target sentences \textbf{\emph{H}} and \textbf{\emph{Z}} and their word embeddings are at hand. A cost means a dissimilarity between $\textbf{\emph{h}}_{i}$ and $\textbf{\emph{z}}_{j}$ (NL and FOL word embeddings) computed by a distance metric $c:\mathbb{R}^d \times \mathbb{R}^d \rightarrow \mathbb{R}_+$, such as cosine distances. The cost matrix $\textbf{\emph{C}} \in \mathbb{R}_+^{m \times n}$ summaries the costs of any word pairs, that is, $\textbf{\emph{C}}_{i,j}=c(\textbf{\emph{h}}_i,\textbf{\emph{z}}_j)$. The OT problem identifies an alignment matrix \emph{P} with which the sum of alignment costs is minimized under the cost matrix \textbf{\emph{C}}:
	\begin{equation}
	L_{\textbf{\emph{C}}(\mu, \nu)} := \min_{\textbf{\emph{P}} \in U(\mu, \nu)} \langle \textbf{\emph{C}}, \textbf{\emph{P}} \rangle,\label{eq2}
	\end{equation}
	where \(U(\mu, \nu) :=  \{ \textbf{\emph{P}} \in \mathbb{R}^{m \times n}_{+} : \textbf{\emph{P}} \mathds{1}_m = \textbf{\emph{a}}, \textbf{\emph{P}}^\top\mathds{1}_n = \textbf{\emph{b}} \}  \).  $\textbf{\emph{P}}_{i,j} \neq 0$ if the \emph{i}-th source word is aligned to the \emph{j}-th target word, such that the aligned words have the smallest distance in the cost matrix $\textbf{\emph{C}}$. With this formulation, we can seek alignment matrix \textbf{\emph{P}}. 
	\section{Methodology}
	\subsection{Overview}\label{Sec4.1}
	The pipeline of our NS-IR is shown in Figure~\ref{fig3}. To reduce the cost of NL-FOL translation, we first use BGE retriever to initially retrieve the top-\emph{K} documents \( D = \{d_1, \ldots, d_i, \ldots, d_K\}\) for a given query of NL (denoted as NL-query). Then, inspired by previous work \cite{olausson-etal-2023-linc, xu-etal-2024-faithful}, we use the specific prompts detailed in Appendix~\ref{prompts} to let an LLM (GPT-4o) perform NL-FOL translation, so as to obtain the query and document of FOL (denoted as FOL-query and FOL-document respectively). Next, we employ BGE\footnote{Significantly, we use BGE twice for different purposes. The first time is to retrieve Top-\emph{K} documents, and the second time is to encode NL and FOL.} to encode the NL-query, FOL-query, NL-document, and FOL-document to obtain corresponding embeddings. Finally, we introduce two independent techniques: logic alignment (Sec.~\ref{sec4.2}) and connective constraint (Sec.,~\ref{sec4.3}) to recalculate the scores between queries and documents and rerank candidate documents\footnote{In this paper, if not specifically stated, queries and documents denote NL-queries and NL-documents, respectively.}.
	
	\begin{table*}[t]
		\centering
		\small
		\begin{tabular}{p{2.5cm}|p{8cm}|c|c|c|c}
			\toprule
			Formulation&Query Example& \#Query&\#Pos.&\#Neg.&\#Irr. \\
			\midrule
			\multirow{2}*{\textbf{A - a}} & Introduce \textcolor{red}{Allen Ginsberg's works}\textsubscript{(A)}, but do not mention '\textcolor{green}{Howl}'\textsubscript{(a)}.&\multirow{2}*{136}&\multirow{2}*{136}&\multirow{2}*{136}&	\multirow{7}*{3000}  \\
			\cline{1-5}
			\multirow{2}*{(\textbf{A - a) $\cup$ B}} & What themes are expressed in \textcolor{red}{Allen Ginsberg's works}\textsubscript{(A)} other than '\textcolor{green}{Howl}'\textsubscript{(a)} and \textcolor{blue}{Edgar Allan Poe's works}\textsubscript{(B)}?&\multirow{2}*{123} &\multirow{2}*{123}&\multirow{2}*{123}\\
			\cline{1-5}
			\multirow{3}*{\textbf{(A - a) $\cup$ (B - b)}} & What themes do  \textcolor{red}{Allen Ginsberg's works}\textsubscript{(A)} other than ' \textcolor{green}{Howl}'\textsubscript{(a)} and \textcolor{blue}{Edgar Allan Poe's works}\textsubscript{(B)} other than '\textcolor{orange}{The Raven}'\textsubscript{(b)} express?&\multirow{3}*{107}&\multirow{3}*{107}&\multirow{3}*{321} \\
			\bottomrule
		\end{tabular}
		\caption{Examples of negative-constraint queries in our constructed dataset NegConstraint. \#Query denotes the number of each type of query, \#Pos., \#Neg., and \#Irr. denote the number of positive documents, and irrelevant documents, respectively.}\label{tab1}
	\end{table*}
	
	\subsection{Logic Alignment}\label{sec4.2}
	To incorporate overall logic semantics in FOL into NL representations, we propose logic alignment based on optimal transport (OT) which is inspired by unsupervised word alignment \cite{arase-etal-2023-unbalanced, huang-etal-2024-ottawa}. This approach measures the distribution differences between NL and FOL, and integrates word features of NL and FOL with context representation via the alignment matrix. 
    
    Specifically, for a given NL-query and FOL-query, we first use BGE to obtain their corresponding word embeddings $\textbf{\emph{H}}$ 
    and $\textbf{\emph{Z}}$\footnote{$\overline{\textbf{\emph{H}}}$ 
    and $\overline{\textbf{\emph{Z}}}$ denote corresponding word embeddings of NL-documents and FOL-documents, respectively.}, respectively.
    Then, we compute a pairwise similarity between \textbf{\emph{H}} and \textbf{\emph{Z}} using cosine distance $\textbf{\emph{C}}_{i,j}$:
	\begin{equation}
	\textbf{\emph{C}}_{i,j} = 1 - \frac{\textbf{\emph{h}}_i^T \textbf{\emph{z}}_j}{\|\textbf{\emph{h}}_i\| \|\textbf{\emph{z}}_j\|}.\label{eq3}
	\end{equation}
	The OT can be formulated as:
	\begin{equation}
	\textbf{\emph{P}}^* = \underset{\textbf{\emph{P}} \in U(\textbf{\emph{H}}, \textbf{\emph{Z}})}{\text{argmin}} \sum_{i,j} \textbf{\emph{C}}_{i,j} \textbf{\emph{P}}_{i,j}, \label{eq4}
	\end{equation}
	where alignment matrix $\textbf{\emph{P}}_{i,j} \in \mathbb{R}^{m \times n}$ indicates a likelihood of aligning $\textbf{\emph{h}}_i$ with $\textbf{\emph{z}}_j$. The optimization in Eq.~\ref{eq4} can be solved by linear programming \cite{bourgeois1971extension}. \\
	\indent Finally, we integrate \textbf{\emph{H}}, \textbf{\emph{Z}}, \textbf{\emph{P}} and $\textbf{\emph{h}}^{cls} \in \mathbb{R}^d$ ($\textbf{\emph{h}}^{cls} \in \mathbb{R}^d$ denote the embeding of special token CLS\footnote{In fact, the first token of queries is denoted as {CLS}, i.e., $\textbf{\emph{h}}^{cls} = \textbf{\emph{h}}_1$.}.) to obtain updated embedding $\mathscr{h}^{cls} \in \mathbb{R}^d$:
	\begin{equation}
	\mathscr{h}^{cls} = \textbf{\emph{H}}^T \cdot \textbf{\emph{P}}  \cdot  \textbf{\emph{Z}} \cdot \textbf{\emph{h}}^{cls}. \label{eq5}
	\end{equation}
	This approach synthesizes word distributions of FOL and NL, as well as context features, via alignment matrix $\textbf{\emph{P}}_{i,j}$ as the intermediary. 
	
	We use the same method to obtain the embedding $\mathscr{d}^{cls} \in \mathbb{R}^d$ of a document $d$ in the document set $D$, and then calculate the similarity score between $\mathscr{h}^{cls}$ and $\mathscr{d}^{cls}$ as
	\begin{equation}
	\text{score}_{1} = \text{sim}\left(\mathscr{h}^{cls} \cdot \mathscr{d}^{cls}\right).\label{eq6}
	\end{equation}
	
	\subsection{Connective Constraint}\label{sec4.3}
	To precisely reflect the role of partial words in FOL on logical consistency, we also propose connective constraint that enables different words in FOL (especially connectives) to render different attentions to words in NL, thus generating better embeddings with logical semantics.  Given a FOL-query sequence \(t = \{t_1, ..., t_i, ..., t_n\}\), as well as NL-query word embeddings \textbf{\emph{H}} and FOL-query word embeddings \textbf{\emph{Z}}, we integrate the embeddings of logical connectives into the attentions. That is, when calculating attention weights (Eq.~\ref{eq8}) of FOL to NL and updated embeddings (Eq.~\ref{eq7}), it takes the alignment between NL and FOL and the embeddings of logical connectives into account:
	\begin{equation}
	\mathscr{h}_j = \sum_{i=1}^{m} \alpha_{ji} \left( \textbf{\emph{h}}_i + \sigma_{ji} \textbf{\emph{z}}_j \right), \label{eq7}
	\end{equation}
	\begin{equation}
	\alpha_{ji} = \frac{e^{\alpha_{ji}^\prime}}{\sum_{i=1}^{m} e^{\alpha_{ji}^\prime}}, \quad \alpha_{ji}^\prime = \frac{\textbf{\emph{z}}_j (\textbf{\emph{h}}_i + \sigma_{ji} \textbf{\emph{z}}_j)^T}{\sqrt{d_k}}, \label{eq8}
	\end{equation}
	\begin{equation}
	\sigma_{ji} = 
	\begin{cases} 
	1, & t_j \neq \neg, t_j \in \mathcal{C}, \textbf{\emph{P}}_{ij} = 0 \\
	-1, & t_j = \neg, \textbf{\emph{P}}_{ij} = 0 \\
	0, & \text{otherwise}
	\end{cases},\label{eq9}
	\end{equation}
	where \textbf{\emph{P}} denotes the alignment matrix and logical connective set \(\mathcal{C} = \{\neg, \rightarrow, \leftrightarrow, \land, \lor, \oplus\}\). 
	
	Our explanations for above equations are as follows.
	For the words that do not exhibit significant alignment between FOL and NL (where \(\textbf{\emph{P}}_{i,j} = 0\)), we incorporate logical connective embeddings to further enhance logical semantics. Additionally, for negative connective (\(t_j = \neg\)), the subtraction implies the negative-constraint semantics.
	
	Finally, we perform mean pooling on $\mathscr{h}_j$ to obtain a query's embedding as
	\begin{equation}
	\mathscr{h} = \text{mean-pooling}(\mathscr{h}_{j}), \quad \mathscr{h} \in \mathbb{R}^d.
	\label{eq10}
	\end{equation}
	Similarly, we obtain the embedding of a document $d$ from the document collection $D$, and then compute the similarity score between $\mathscr{h}$ and $\mathscr{d}$ as
	\begin{equation}
	\text{score}_2 = \text{sim}(\mathscr{h} \cdot \mathscr{d}).\label{eq11}
	\end{equation}
	\indent For document $d$, the final recommended score is
	\begin{equation}
	\text{score}= \text{score}_1 + \text{score}_2. \label{eq12}
	\end{equation}
	Then, we rerank the top-\emph{K} candidate documents based on the recommendation scores.
    
\begin{table*}
\centering	
	\resizebox{1.02\textwidth}{!}{		
		\begin{tabular}{l|cccccc|cc|cc}
			\toprule
			Methods & SciFact & ArguAna & \small{TREC-COVID} & FiQA & DBPedia & NFCorpus& \multicolumn{2}{c|}{DL'19} & \multicolumn{2}{c}{DL'20} \\
            \midrule
            \rowcolor[rgb]{ .9,  .9,  .9} 
			\textbf{w/ relevance judgment} & \multicolumn{6}{c|}{nDCG@10}& MAP & \small{nDCG@10} & MAP & \small{nDCG@10} \\
			DPR  & 31.8 & 17.5 & 33.2 & 29.5 & 26.3 & 18.9& 36.5 & 62.2  & 41.8 & 65.3  \\
			ANCE & 50.7 & 41.5 & 65.4 & 30.0 & 28.1 & 23.7& 37.1 & 64.5  & 40.8 & 64.6  \\
			Contriever$^{\text{FT}}$ & 67.7 & 44.6 & 59.6 & 32.9 & 41.3 & 32.8 & 41.7 & 62.1 & 43.6 & 63.2  \\
			\midrule
            \rowcolor[rgb]{ .9,  .9,  .9} 
			\textbf{w/o relevance judgment} & \multicolumn{6}{c|}{nDCG@10}& MAP & \small{nDCG@10} & MAP & \small{nDCG@10} \\
			BM25$\heartsuit$ & 67.1 & 43.2 & 55.5 & 25.1 & 26.1 & 31.4& 31.2 & 55.4  & 30.6 & 50.1  \\
			Contriever$\heartsuit$& 55.0 &44.5 & 12.5  & 12.4 & 29.2 & 26.0 & 22.8 & 37.5 & 24.3 & 42.5\\
			HyDE$\heartsuit$ & 71.9 & 49.6 & 78.4 & 31.3 & 38.7 & 37.3& 48.7 & 67.3  & 49.8 & 66.8 \\
			InterR$\heartsuit$& 72.1& 50.9 & 79.2 & 33.5 &42.1 &39.5 & 50.4 & \textbf{69.7}& 47.8 & 67.5\\
			BGE$\heartsuit^*$ & 71.3 &  48.4& 75.3 & 30.6 & 38.9 & 35.4& 46.9 & 64.4 & 45.7 & 63.4 \\
			BGE w/ LA$\heartsuit^*$ &72.6 &53.2 &78.3 &33.7 &42.8&38.1&48.9 &67.5  &47.5 &68.9 \\
			BGE w/ CC$\heartsuit^*$ &73.3 &52.3 &77.6 &35.6 &43.2 &37.7&49.1 &66.8 &48.5 &66.6 \\
			\midrule
			NS-IR (LogicLLaMA)&73.7 &51.1 &78.8 &35.5 &42.6 &38.8& 49.8& 67.9  & 48.9 & 68.1  \\
            \rowcolor[rgb]{ .9,  .9,  .9} 
			NS-IR (GPT-4o)& \textbf{75.8} & \textbf{55.1} & \textbf{81.8} &\textbf{38.4} &\textbf{ 46.1} &\textbf{40.7}& \textbf{51.4 }& 68.4 & \textbf{50.8} & \textbf{70.5} \\
			\bottomrule
		\end{tabular}
        }
		\caption{Performance of compared methods on the benchmarks of low-resource retrieval and web search.  $\heartsuit$ indicates the reported results were reproduced by us using the baselines' sourcecodes. We employ BGE as the embedding model in HyDE and InteR for fair comparison. 
			$^*$ denotes the ablated variants of NS-IR which can be regarded as BGE w/ LA\&CC. }\label{tab2}
	\end{table*}
    
	\section{NegConstraint}\label{sec5}
	To evaluate our NS-IR's performance specifically to negative-constraint queries, we have constructed a human-annotated dataset \textbf{NegConstraint}. Table~\ref{tab1} lists three formulations that represent three types of negative-constraints query. Let the letters (A, a, B, and b) denote entity sets where the lowercase letters represent the subsets of the sets denoted by the uppercase letters. Operator `-' indicates a negative-constraint condition, and `$\cup$' denotes union operation. There are 136 queries corresponding to the first formulation, 123 queries corresponding to the second formulation, and 107 queries corresponding to the third formulation in NegConstraint. 
	Each query is paired with one positive document and one or three negative documents as distractors. For example, for a query of formulation \textbf{A – a}, such as ``Introduce Allen Ginsberg's works, but do not mention `Howl''', there is a positive document that introduces the content about Allen Ginsberg's works but does not mention `Howl'. Similarly, a negative document introduces Allen Ginsberg's and mentions `Howl'. The documents stem from the Wikipedia dump \cite{karpukhin-etal-2020-dense}, and the queries are generated by GPT-4o based on corresponding negative and positive documents where the prompts are presented in Appendix~\ref{prompts}. Besides, NegConstraint contains 3,000 irrelevant documents\footnote{Negative documents are essentially irrelevant documents, but easier to mislead retriever models.} that are irrelevant to all queries. More details about our data collection and snippets are provided in Appendix~\ref{negconstraint}.

\section{Experiments}
	\subsection{Datasets and Metrics}
	For \textbf{low-resource retrieval}, we use six diverse low-resource retrieval datasets from the BEIR benchmark \cite{thakur2021heterogenous}, including SciFact (fact-checking), ArguAna (argument retrieval), TREC-COVID (bio-medical IR), FiQA (financial QA), DBPedia (entity retrieval), and NFCorpus (medical IR). For this task, we report all compared methods' scores of the representative metric nDCG@10. For \textbf{web search}, we adopt widely-used web search dataset TREC Deep Learning 2019 (DL’19) \cite{craswell2020overview} and Deep Learning 2020 (DL’20) \cite{craswell2021overviewtrec2020deep} based on MS-MARCO \cite{bajaj2018human}. For these two datasets and our NegConstraint, we report the methods' scores of MAP (Mean Average Precision) and nDCG@10.

    \begin{table*}[h]
		\small
		\centering
		\begin{tabular}{l|cc|cc|cc|cc}
			\toprule
			\multirow{2.5}{*}{Methods}
			& \multicolumn{2}{c|}{\textbf{A -  a}} & \multicolumn{2}{c|}{\textbf{(A -  a) $\cup$ B}}& \multicolumn{2}{c|}{\textbf{(A -  a) $\cup$ (B - b)}}& \multicolumn{2}{c}{Total} \\
			\cmidrule{2-3}\cmidrule{4-5}\cmidrule{6-7}\cmidrule{8-9}
			& MAP & nDCG@10& MAP & nDCG@10& MAP & nDCG@10& MAP & nDCG@10\\
			\midrule
			BM25$\heartsuit$ & 32.1 & 34.6 & 31.2 & 34.7 & 29.3 & 31.5& 31.4 & 33.7 \\
			Contriever$\heartsuit$ & 34.8 & 36.6 & 32.1 & 33.3 & 30.9 & 32.7& 31.8 & 35.7 \\
			HyDE$\heartsuit$ & 50.7 & 55.3 & 48.6 & 51.5 &  45.7& 50.6& 47.8 & 53.1 \\
			InterR$\heartsuit$ & 52.6 & 55.8 &50.3& 48.7 & 51.5 & 49.3& 52.3 & 54.5\\
			BGE$\heartsuit^*$ & 37.9 & 40.5 & 34.8 & 36.8 & 33.7 & 34.8& 36.3 &40.8 \\
			BGE w/ LA$\heartsuit^*$ & 42.1 & 45.6 & 41.2 & 44.6 & 39.9 & 42.5& 40.8 &47.6 \\
			BGE w/ CC$\heartsuit^*$ & 48.9 & 50.7 & 46.6 & 48.5 & 43.9 & 48.2& 47.8 &46.9 \\
			\midrule
			NS-IR (LogicLLaMA)&53.2& 54.6 & 51.6 &50.2 & 49.6 & \textbf{54.1}& 50.7 & 55.2 \\
            \rowcolor[rgb]{ .9,  .9,  .9} 
			NS-IR (GPT-4o)&\textbf{54.7}& \textbf{57.9} & \textbf{53.3} & \textbf{54.2} & \textbf{51.7} & 53.7& \textbf{53.3} & \textbf{56.5} \\
			\bottomrule
		\end{tabular}
		\caption{Performance comparisons of different methods on NegConstraint.}
		\label{tab4}
	\end{table*}
    
	\subsection{Baselines}
    We first compare our NS-IR with some fully-supervised retrieval methods that are fine-tuned with extensive query-document relevance data (denoted as \textbf{w/ relevance judgment}), including DPR \cite{karpukhin2020dense}, ANCE \cite{xiong2020approximate}, and the fine-tuned Contriever (denoted as Contriever$^{FT}$ \cite{izacard2021unsupervised}).
    We also consider several zero-shot retrieval models not involving query-document relevance labels (denoted as \textbf{w/o relevance judgment}), including sparse retriever BM25 \cite{robertson2009probabilistic}, dense retriever BGE \cite{xiao2024cpackpackedresourcesgeneral}, and Contriever \cite{izacard2021unsupervised}. For this type of baselines, we further consider the LLM-based retrieval models which rewrite queries with LLMs, including HyDE \cite{gao-etal-2023-precise} and InteR \cite{feng-etal-2024-synergistic}. 
    \subsection{Implementation Details} \label{MethodsandImplementation}
    We employ "bge-large-en-v1.5" as embedding models. To make a fair comparison, we also reproduce results on HyDE and InteR where "bge-large-en-v1.5" is used as retriever models. We run all experiments on one Nvidia A800 80GB GPU. For NL-FOL translation, we use OpenAI API with a temperature of 0.5.
    
	\subsection{Main Results}
    In the following presentation, the techniques of logic alignment and connective constraint we proposed for NS-IR are abbreviated as LA and CC, respectively. In our comparison experiments, we adopt two variants of NS-IR which use LogicLLaMA \cite{yang-etal-2024-harnessing} and GPT-4o to generate FOL, respectively.
	
    Table~\ref{tab2} shows that, our NS-IR (GPT-4o) outperforms all baselines significantly on the tasks of low-resource retrieval and web search, including the SOTA model without relevance judgment InteR. 
    Specifically, NS-IR (GPT-4o) obtains an average performance improvement of over 10\% relative to the vanilla BGE. 
    The inferiority of NS-IR (LogicLLaMA) compared to NS-IR (GPT-4o) is attributed to LogicLLaMA's weakness on NL-FOL translation.
    
    
    For negative-constraint queries, we compare NS-IR and its ablated variants with the baselines without relevance judgment.
    Table~\ref{tab4} reports the compared methods' performance on three types of negative-constraint queries and whole queries in NegConstraint. 
    The results reveal our method's superiority over the baselines
    on negative-constraint queries, 
    which is achieved through synthesizing semantic similarity and logical consistency for handling complex logical queries.
    Although HyDE and InterR can partially eliminate the impact of negative constraints via hypothetical documents generated by LLMs, our proposed LA and CC are more effective. 
	

    The results in Tables~\ref{tab2} and \ref{tab4} related to NS-IR's ablated variants (marked by $^*$) also justify the effectiveness of employing either LA or CC. 
    In particular, adopting CC improves NS-IR's performance more obviously than adopting LA on NegConstraint, suggesting that CC is more effective than LA in the scenarios of negative-constraint queries.
	
	\begin{table}[t]
		\centering
		\small
		\begin{tabular}{l|cc|cc}
			\toprule
			\multirow{2.5}{*}{Methods} & \multicolumn{2}{c|}{DL'19}   & \multicolumn{2}{c}{DL'20} \\
			\cmidrule{2-3}\cmidrule{4-5}
			& MAP & nDCG@10 & MAP & nDCG@10 \\
			\midrule
			bge-small & 40.5 & 60.1 & 40.4 & 61.2 \\
			\quad + HyDE & 42.7 & 61.4 & 41.8 & 62.7 \\
			\quad + InteR & 43.6 & 63.4 & 42.8 & 63.2 \\
            \rowcolor[rgb]{ .9,  .9,  .9} 
			\quad + NS-IR & \textbf{43.8} & \textbf{65.4} & \textbf{44.4} & \textbf{64.9} \\
			\midrule
			bge-base & 41.9 & 62.5 & 40.9 & 63.1 \\
			\quad + HyDE & 42.4 & 64.4 & 42.6 & 64.5 \\
			\quad + InteR & 45.9 & 65.3 & 45.5 & 66.7 \\
            \rowcolor[rgb]{ .9,  .9,  .9} 
			\quad + NS-IR & \textbf{46.6} & \textbf{66.4} & \textbf{47.9} & \textbf{68.8} \\
			\midrule
			Contriever & 35.7 & 57.7 & 37.8 & 56.9 \\
			\quad + HyDE & 37.5 & 59.3 & 39.9 & 58.9 \\
			\quad + InteR & 38.6 & 58.9 & 39.8 & 61.7 \\
            \rowcolor[rgb]{ .9,  .9,  .9} 
			\quad + NS-IR & \textbf{40.6} & \textbf{61.4} & \textbf{41.4} & \textbf{62.8} \\
			\bottomrule
		\end{tabular}
		\caption{Web search performance of adopting different dense retriever models in HyDE, InteR and our NS-IR.}\label{tab6}
	\end{table}
    
	\begin{figure*}[h]
		\centering
		\includegraphics[width=1\linewidth]{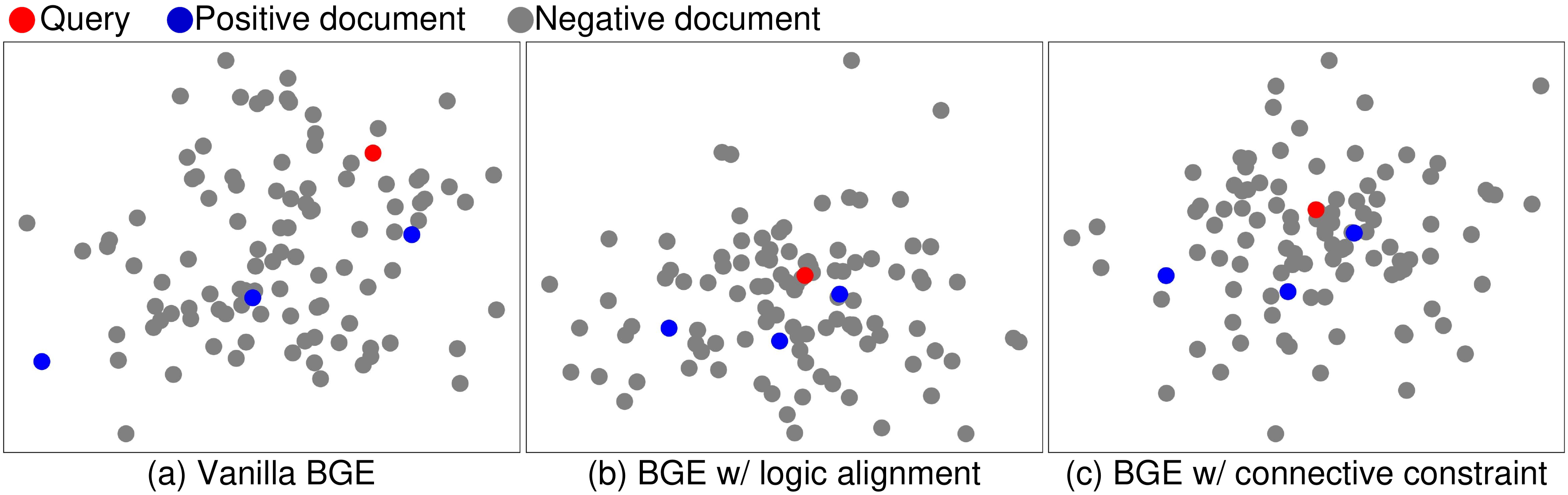}
		\caption{An example of query embedding visualization from TREC-COVID (better viewed in color): \emph{What are the observed mutations in the SARS-CoV-2 genome and how often do the mutations occur?}}\label{fig5}
	\end{figure*}
	\begin{figure*}[h]
		\centering
		\includegraphics[width=1\linewidth]{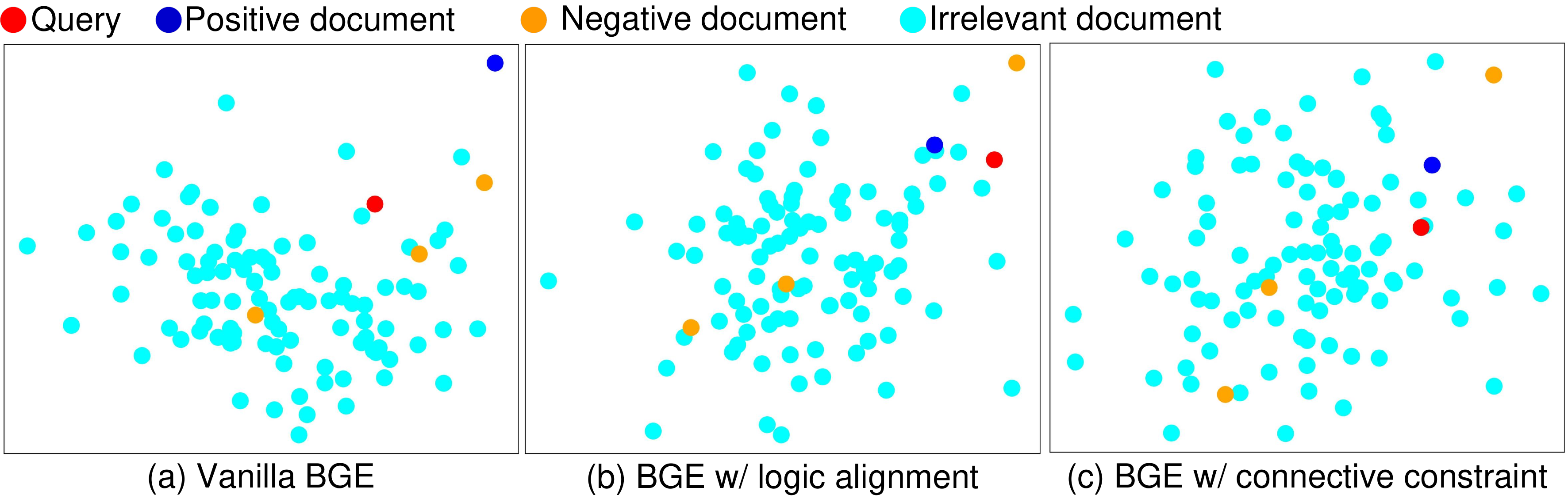}
		\caption{An example of query embedding visualization from NegConstraint (better viewed in color): \emph{What are the similarities between Ginsberg's works (excluding 'Howl') and Poe's works (excluding 'The Raven')?}}\label{fig6}
	\end{figure*}
    
    \subsection{Effects of Different Dense Retrievers}
	To verify the effectiveness of different dense retrievers, we report the web search performance of HyDE, InteR, and NS-IR (GPT-4o) with different dense retrievers (bge-small, bge-base, and Contriver)\footnote{We replace dense retrievers in the process of retrieval and encoding as introduced in Sec.~\ref{Sec4.1}.} in Table~\ref{tab6}. The results indicate that more powerful retriever models can facilitate accurate IR. NS-IR is consistently superior to HyDE and InteR with all retrievers. These results also indicate that the effectiveness of LA and CC on NS-IR's performance gains are model-agnostic.
	\subsection{Visualization on the Effects of Logic Alignment and Connective Constraint}
	We randomly pick two queries from TREC-COVID and our NegConstraint to visualize the effects of LA and CC. In Figures~\ref{fig5} and~\ref{fig6}, we plot the embeddings generated by vanilla BGE, BGE w/ LA and BGE w/ CC in the embedding space using t-SNE, respectively. In Figure~\ref{fig5} of TREC-COVID, we can see that the query embeddings generated by BGE w/ LA and BGE w/ CC are closer to that of positive documents than the query embeddings generated by vanilla BGE. In Figure~\ref{fig6} of NegConstraint, the query embeddings generated by BGE w/ LA and BGE w/ CC are closer to that of positive documents and farther away from that of negative documents, compared to the query embeddings generated by vanilla BGE. 
    These results demonstrate that LA anc CC are more effective on identifying positive documents.
	\subsection{Visualization on the Attention of Connective Constraint}
	\begin{figure}[t]
		\centering
		\includegraphics[width=1.0\linewidth]{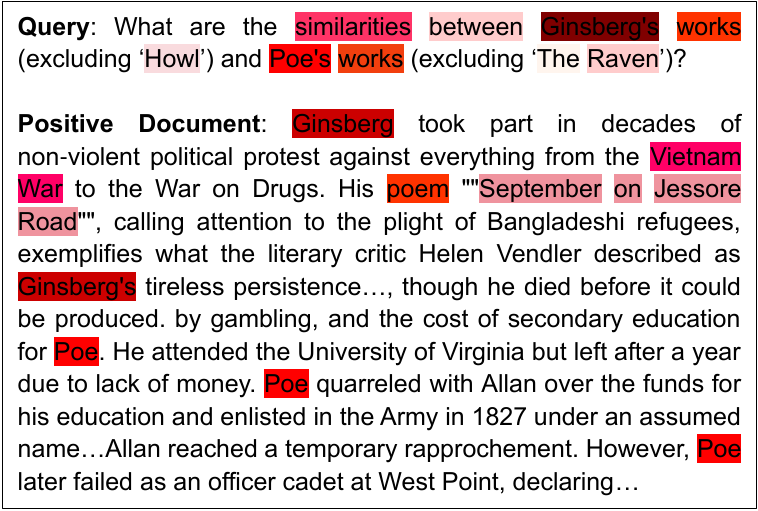}
		\caption{Attention scores of logical connective $\neg$ in FOL to the words in NL. A deeper color indicates a bigger score (better viewed in color).}\label{fig7}
	\end{figure}
	As introduced in Sec.~\ref{sec4.2}, CC enables the words in FOL (especially connectives) to assign different attentions to different words in NL. To verify the hypothesis, we examine the attention scores of logical negation $\neg$ in FOL to the words in NL. As shown in Figure~\ref{fig7}, the deeper colors indicate the larger attention scores. We suppose that this operation emphasizes important entity tokens (such as `Ginsberg' and `Poe') and ignores entity tokens in negative-constraint conditions (such as `Howl' and `Raven'). That is, logical connective $\neg$ implies negative-constraint semantics. It reveals that our method tends to retrieve the documents without negative-constraint conditions mentioned in queries. 
    \section{Effect of Parameter \emph{K}}
	\begin{figure}[h]
		\centering
		\includegraphics[width=\linewidth]{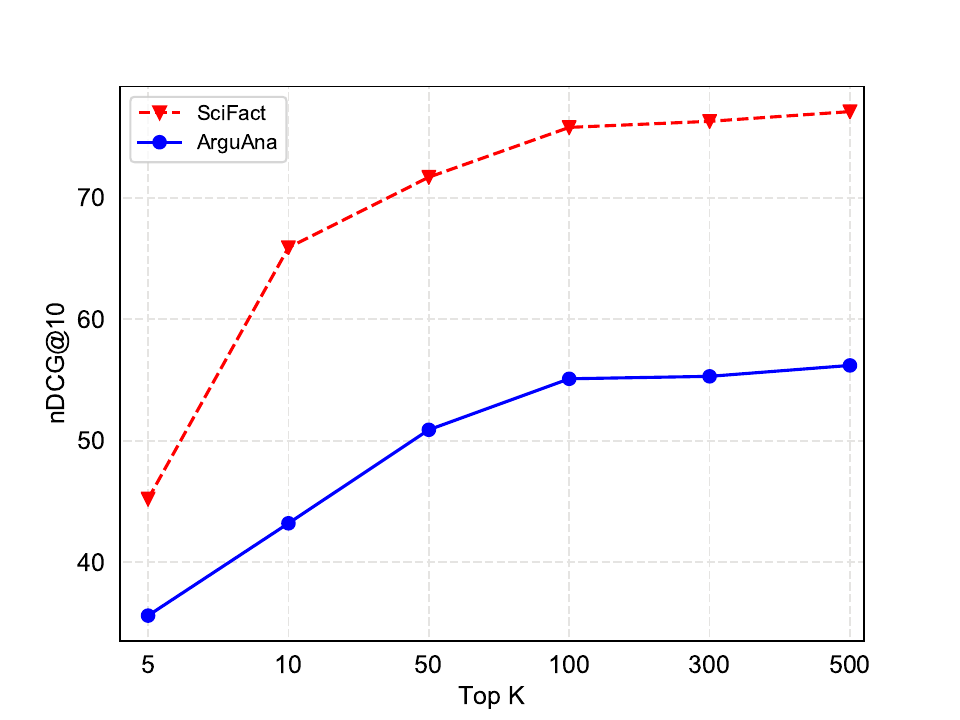}
		\caption{The performance of NS-IR on different Top \emph{K} of SciFact and ArguAna.}\label{fig4}
	\end{figure}
	We perform an additional study to investigate the impact of the number of retrieved documents (i.e., Top \emph{K}) on the performance of NS-IR. Figure~\ref{fig4} illustrates nDCG@10 under different \emph{K} on SciFact and ArguAna. Our observations revealed consistent patterns in both datasets: as Top \emph{K} increased, performance showed a gradual improvement until \emph{K} reached 100. Subsequently, the performance stabilized, indicating that increased \emph{K} did not significantly enhance the results. This phenomenon can be attributed to the fact that the first 100 retrieved documents have already covered a significant amount of positive documents for a query. Additionally, the larger number of retrieval documents will extremely increase expenses for generating FOL. Therefore, we select 100 as Top \emph{K} in this paper.
	\section{Conclusion}
	In this paper, we propose a novel IR method NS-IR, which integrates the strengths of NL and FOL and synthesizes semantic similarity and logical consistency. We specially propose two key techniques: logic alignment and connective constraint, to rerank the candidate documents. We also release a negative-constraint query dataset NegConstraint to evaluate our method. Extensive experiments on public IR benchmarks and NegConstraint show that, NS-IR significantly outperforms the existing IR approaches for general and negative-constraint queries under zero-shot settings, paving the way for future study on complex logical queries. Therefore, we will focus on more complex logical queries generated by set operations (such as union, intersection, difference, and complement) in the future.
	\section{Acknowledgments}
      The authors disclosed receipt of the following financial support for the research, authorship, and publication of this article: This research was supported by the AI Laboratory of United Automotive Electronic Systems (UAES) Co. (Grant no. 2025-3944) and the Chinese NSF Major Research Plan (No.92270121). 
	\section*{Limitations}
	We acknowledge that our method has several limitations. First, calling OpenAI API to perform NL-FOL translation will inevitably incur additional expenses to maintain high retrieval relevance.  Second, to reduce the expenses of NL-FOL translation, we perform NL-FOL translation on the initially retrieved and limited documents, thus slightly reducing recall. Third, we use the same prompts for NL-FOL translation on all benchmarks, which may hinder further improvement. Therefore, these limitations are caused by NL-FOL translation. Although NL-FOL translation is not the main focus of this paper, we argue that the limitations will be improved with further study in the era of LLMs.
	
	\bibliography{arxiv}
	\appendix
	\begin{table*}
		\centering
		\small
		\begin{tabular}{c|p{13cm}}
			\toprule
			NL-query &What is considered a business expense on a business trip? \\
			\midrule
			FOL-query &$\forall$x (BusinessTrip(x) $\rightarrow$ BusinessExpense(x)) \\
			\midrule
			\multirow{5}*{NL-documnet}&I'm not saying I don't like the idea of on-the-job training too, but you can't expect the company to do that. Training workers is not their job - they're building software. Perhaps educational systems in the U.S. (or their students) should worry a little about getting marketable skills in exchange for their massive investment in education, rather than getting out with thousands in student debt and then complaining that they aren't qualified to do anything.\\
			\midrule
			FOL-document& \makecell{$\neg$Like(i, onTheJobTraining) $\land$$\neg$Expect(company, Train(workers))  \\$\neg$Job(company, Train(workers))\\ $\land$Job(company, Build(software)) $\forall$x (EducationalSystems(x) $\rightarrow$ Worry(x, MarketableSkills)) \\Invest(educationalSystems, education) $\land$$\exists$x (Student(x) $\land$StudentDebt(x)) $\rightarrow$ $\neg$Qualified(x, anything)}\\
			\bottomrule
		\end{tabular}
		\caption{An example of the NL-query, FOL-query, NL-document and FOL-document.}\label{tab7}
	\end{table*}
	\section{FOL Examples}\label{folexamples}
	Table~\ref{tab7} provides an example of the NL-query, FOL-query, NL-document, and FOL-document. The NL-query and NL-document stem from actual datasets. FOL-query and FOL-document are FOL formats of NL-queries and NL-documents, respectively, which are generated by GPT-4o in this paper.
    \onecolumn
	\section{Prompts}\label{prompts}
	\begin{tcolorbox}[enhanced, colback=white, colframe=blue!50!black, title=Prompt of NL-FOL Translation for Queries, fonttitle=\bfseries, fontupper=\small]
		Given some question. The task is to parse these questions into first-order logic formulars. The grammar of the first-order logic formular is defined as follows:\\
		1) logical conjunction of expr1 and expr2: expr1 $\land$ expr2\\
		2) logical disjunction of expr1 and expr2: expr1 $\lor$ expr2\\
		3) logical exclusive disjunction of expr1 and expr2: expr1 $\oplus$ expr2\\
		4) logical negation of expr1: $\neg$expr1\\
		5) expr1 implies expr2: expr1 $\rightarrow$ expr2\\
		6) expr1 if and only if expr2: expr1 $\leftrightarrow$ expr2\\
		7) logical universal quantification: $\exists$x\\
		8) logical existential quantification: $\forall$x\\
		------\\
		Here is an example:\\
		Query:\\
		Rina is either a person who jokes about being addicted to caffeine or is unaware that caffeine is a drug.\\
		If Rina is either a person who jokes about being addicted to caffeine and a person who is unaware that caffeine is a drug, or neither a person who jokes about being addicted to caffeine nor a person who is unaware that caffeine is a drug, then Rina jokes about being addicted to caffeine and regularly drinks coffee.\\
		\#\#\#\\
		Predicates:\\
		Drinks(x) ::: x regularly drinks coffee.\\
		Jokes(x) ::: x jokes about being addicted to caffeine.\\
		Unaware(x) ::: x is unaware that caffeine is a drug.\\
		Conclusion:\\
		Jokes(rina) $\oplus$ Unaware(rina) ::: Rina is either a person who jokes about being addicted to caffeine or is unaware that caffeine is a drug.\\
		((Jokes(rina) $\land$ Unaware(rina))  $\oplus$  $\neg$(Jokes(rina) $\lor$ Unaware(rina))) $\rightarrow$ (Jokes(rina) $\land$ Drinks(rina)) ::: If Rina is either a person who jokes about being addicted to caffeine and a person who is unaware that caffeine is a drug, or neither a person who jokes about being addicted to caffeine nor a person who is unaware that caffeine is a drug, then Rina jokes about being addicted to caffeine and regularly drinks coffee.\\
		------\\
		Here is an example:\\
		Query:\\
		Miroslav Venhoda loved music.\\
		A Czech person wrote a book in 1946.\\
		No choral conductor specialized in the performance of Renaissance.\\
		\#\#\#\\
		Predicates:\\
		Czech(x) ::: x is a Czech person.\\
		ChoralConductor(x) ::: x is a choral conductor.\\
		Author(x, y) ::: x is the author of y.\\
		Book(x) ::: x is a book.\\
		Specialize(x, y) ::: x specializes in y.\\
		Conclusion:\\
		Love(miroslav, music) ::: Miroslav Venhoda loved music.\\
		$\exists$y $\exists$x (Czech(x) $\land$ Author(x, y) $\land$ Book(y) $\land$ Publish(y, year1946)) ::: A Czech person wrote a book in 1946.\\
		$\neg$ $\exists$x (ChoralConductor(x) $\land$ Specialize(x, renaissance)) ::: No choral conductor specialized in the performance of Renaissance.\\
		------\\
		Below is the one you need to translate:\\
		Query:\\
		\%QUERY\% \\
		-------
	\end{tcolorbox}
	\begin{tcolorbox}[enhanced, colback=white, colframe=blue!50!black, title=Prompt of NL-FOL Translation for Documents, fonttitle=\bfseries, fontupper=\small]
		Given a document. The task is to parse the document into first-order logic formulars. The grammar of the first-order logic formular is defined as follows:\\
		1) logical conjunction of expr1 and expr2: expr1 $\land$ expr2\\
		2) logical disjunction of expr1 and expr2: expr1 $\lor$ expr2\\
		3) logical exclusive disjunction of expr1 and expr2: expr1 $\oplus$ expr2\\
		4) logical negation of expr1: $\neg$expr1\\
		5) expr1 implies expr2: expr1 $\rightarrow$ expr2\\
		6) expr1 if and only if expr2: expr1 $\leftrightarrow$ expr2\\
		7) logical universal quantification: $\exists$x\\
		8) logical existential quantification: $\forall$x\\
		------\\
		Here is an example:\\
		Document:\\
		All people who regularly drink coffee are dependent on caffeine. People either regularly drink coffee or joke about being addicted to caffeine. No one who jokes about being addicted to caffeine is unaware that caffeine is a drug. Rina is either a student and unaware that caffeine is a drug, or neither a student nor unaware that caffeine is a drug. If Rina is not a person dependent on caffeine and a student, then Rina is either a person dependent on caffeine and a student, or neither a person dependent on caffeine nor a student.\\
		\#\#\#\\
		Predicates:\\
		Dependent(x) ::: x is a person dependent on caffeine.\\
		Drinks(x) ::: x regularly drinks coffee.\\
		Jokes(x) ::: x jokes about being addicted to caffeine.\\
		Unaware(x) ::: x is unaware that caffeine is a drug.\\
		Student(x) ::: x is a student.\\
		Conclusion:\\
		$\forall$x (Drinks(x) $\rightarrow$ Dependent(x)) ::: All people who regularly drink coffee are dependent on caffeine.\\
		$\forall$x (Drinks(x) $\oplus$ Jokes(x)) ::: People either regularly drink coffee or joke about being addicted to caffeine.\\
		$\forall$x (Jokes(x) $\rightarrow$ $\neg$Unaware(x)) ::: No one who jokes about being addicted to caffeine is unaware that caffeine is a drug.\\
		(Student(rina) $\land$ Unaware(rina)) $\oplus$ $\neg$(Student(rina) $\lor$ Unaware(rina)) ::: Rina is either a student and unaware that caffeine is a drug, or neither a student nor unaware that caffeine is a drug.\\
		$\neg$(Dependent(rina) $\land$ Student(rina)) $\rightarrow$ (Dependent(rina) $\land$ Student(rina)) $\oplus$ $\neg$(Dependent(rina) $\lor$ Student(rina)) ::: If Rina is not a person dependent on caffeine and a student, then Rina is either a person dependent on caffeine and a student, or neither a person dependent on caffeine nor a student.\\
		------\\
		Here is an example:\\
		Document:\\
		Miroslav Venhoda was a Czech choral conductor who specialized in the performance of Renaissance and Baroque music. Any choral conductor is a musician. Some musicians love music. Miroslav Venhoda published a book in 1946 called Method of Studying Gregorian Chant.\\
		\#\#\#\\
		Predicates:\\
		Czech(x) ::: x is a Czech person.\\
		ChoralConductor(x) ::: x is a choral conductor.\\
		Musician(x) ::: x is a musician.\\
		Love(x, y) ::: x loves y.\\
		Author(x, y) ::: x is the author of y.\\
		Book(x) ::: x is a book.\\
		Publish(x, y) ::: x is published in year y.\\
		Specialize(x, y) ::: x specializes in y.\\
		Conclusion:\\
		Czech(miroslav) $\land$ ChoralConductor(miroslav) $\land$ Specialize(miroslav, renaissance) $\land$ Specialize(miroslav, baroque) ::: Miroslav Venhoda was a Czech choral conductor who specialized in the performance of Renaissance and Baroque music.\\
		$\exists$x (ChoralConductor(x) $\rightarrow$ Musician(x)) ::: Any choral conductor is a musician.\\
		$\forall$x (Musician(x) $\land$ Love(x, music)) ::: Some musicians love music.\\
		Book(methodOfStudyingGregorianChant) $\land$ Author(miroslav, methodOfStudyingGregorianChant) $\land$ Publish(methodOfStudyingGregorianChant, year1946) ::: Miroslav Venhoda published a book in 1946 called Method of Studying Gregorian Chant.\\
		------\\
		Below is the one you need to translate:\\
		Document:\\
		\%DOCUMENT\%
	\end{tcolorbox}
	\begin{tcolorbox}[enhanced, colback=white, colframe=blue!50!black, title=Prompt of Query Generation for \textbf{A - a}, fonttitle=\bfseries, fontupper=\small]
		Given an example in information retrieval tasks. We refer to the query as a negative-constraint query. The query matches formulation \textbf{A - a}. \textbf{A} denotes Ginsberg's works and \textbf{a} denotes 'Howl'. The positive document mentions Ginsberg's works but does not mention 'Howl'. The negative document mentions Ginsberg's works and 'Howl'. Please provide a query based on the positive and negative documents provided.\\
		------\\
		\textbf{EXAMPLE}\\
		\textbf{Positive document}:\\
		Ginsberg took part in decades of non-violent political protest against everything from the Vietnam War to the War on Drugs. His poem ""September on Jessore Road"", calling attention to the plight of Bangladeshi refugees, exemplifies what the literary critic Helen Vendler described as Ginsberg's tireless persistence in protesting against ""imperial politics, and persecution of the powerless."" His collection ""The Fall of America"" shared the annual U.S. National Book Award for Poetry in 1974. In 1979 he received the National Arts Club gold medal and was inducted into the American Academy and Institute of Arts and Letters. Ginsberg was a Pulitzer. by gambling, and the cost of secondary education for Poe. He attended the University of Virginia but left after a year due to lack of money. Poe quarreled with Allan over the funds for his education and enlisted in the Army in 1827 under an assumed name. It was at this time that his publishing career began, albeit humbly, with the anonymous collection ""Tamerlane and Other Poems"" (1827), credited only to ""a Bostonian"". With the death of Frances Allan in 1829, Poe and Allan reached a temporary rapprochement. However, Poe later failed as an officer cadet at West Point, declaring...\\
		\textbf{Negative document}: \\
		Kerouac and William S. Burroughs. Ginsberg is best known for his poem ""Howl"", in which he denounced what he saw as the destructive forces of capitalism and conformity in the United States. In 1956, ""Howl"" was seized by San Francisco police and US Customs. In 1957, it attracted widespread publicity when it became the subject of an obscenity trial, as it described heterosexual and homosexual sex at a time when sodomy laws made homosexual acts a crime in every U.S. state. ""Howl"" reflected Ginsberg's own homosexuality and his relationships with a number of men, including Peter Orlovsky, his lifelong partner.\\
		\textbf{Query}:\\
		Introduce Allen Ginsberg's works, but do not mention 'Howl'.\\
		------\\
		\textbf{Positive document}: \%POSITIVE DOCUMENT\%\\
		\textbf{Negative document}: \%NEGATIVE DOCUMENT\%\\
		------\\
		Below query is the one you need to generate, which make significant changes to the query style.\\
		\textbf{Query}:\\
		\%QUERY\%
	\end{tcolorbox}
	\begin{tcolorbox}[enhanced, colback=white, colframe=blue!50!black, title=Prompt of Query Generation \textbf{(A - a) $\cup$ B}, fonttitle=\bfseries, fontupper=\small]
		Given an example in information retrieval tasks. We refer to the query as a negative-constraint query. The query matches formulation \textbf{(A - a) $\cup$ B}. \textbf{A} denotes Allen Ginsberg's works, \textbf{a} denotes 'Howl', and \textbf{B} denotes Edgar Allan Poe's works. The positive document mentions Allen Ginsberg's and Edgar Allan Poe's works but does not mention 'Howl'. The negative document mentions Allen Ginsberg's works,  Edgar Allan Poe's works and 'Howl'. Please provide a query based on the positive and negative documents provided.\\
		------\\
		\textbf{EXAMPLE}\\
		\textbf{Positive document}:\\
		Ginsberg took part in decades of non-violent political protest against everything from the Vietnam War to the War on Drugs. His poem ""September on Jessore Road"", calling attention to the plight of Bangladeshi refugees, exemplifies what the literary critic Helen Vendler described as Ginsberg's tireless persistence in protesting against ""imperial politics, and persecution of the powerless."" His collection ""The Fall of America"" shared the annual U.S. National Book Award for Poetry in 1974. In 1979 he received the National Arts Club gold medal and was inducted into the American Academy and Institute of Arts and Letters. Ginsberg was a Pulitzer. by gambling, and the cost of secondary education for Poe. He attended the University of Virginia but left after a year due to lack of money. Poe quarreled with Allan over the funds for his education and enlisted in the Army in 1827 under an assumed name. It was at this time that his publishing career began, albeit humbly, with the anonymous collection ""Tamerlane and Other Poems"" (1827), credited only to ""a Bostonian"". With the death of Frances Allan in 1829, Poe and Allan reached a temporary rapprochement. However, Poe later failed as an officer cadet at West Point, declaring...\\
		\textbf{Negative document}: \\
		Kerouac and William S. Burroughs. Ginsberg is best known for his poem ""Howl"", in which he denounced what he saw as the destructive forces of capitalism and conformity in the United States. In 1956, ""Howl"" was seized by San Francisco police and US Customs. In 1957, it attracted widespread publicity when it became the subject of an obscenity trial, as it described heterosexual and homosexual sex at a time when sodomy laws made homosexual acts a crime in every U.S. state. ""Howl"" reflected Ginsberg's own homosexuality and his relationships with a number of men, including Peter Orlovsky, his lifelong partner. by gambling, and the cost of secondary education for Poe. He attended the University of Virginia but left after a year due to lack of money. Poe quarreled with Allan over the funds for his education and enlisted in the Army in 1827 under an assumed name. It was at this time that his publishing career began, albeit humbly, with the anonymous collection ""Tamerlane and Other Poems"" (1827), credited only to ""a Bostonian"". With the death of Frances Allan in 1829, Poe and Allan reached a temporary rapprochement. However, Poe later failed as an officer cadet at West Point, declaring...\\
		\textbf{Query}:\\
		What themes are expressed in Allen Ginsberg's works (excluding "Howl")? Is there any similarity between Edgar Allan Poe's works and theirs?\\
		------\\
		\textbf{Positive document}: \%POSITIVE DOCUMENT\%\\
		\textbf{Negative document}: \%NEGATIVE DOCUMENT\%\\
		------\\
		Below query is the one you need to generate, which make significant changes to the query style.\\
		\textbf{Query}:\\
		\%QUERY\%
	\end{tcolorbox}
	\begin{tcolorbox}[enhanced, colback=white, colframe=blue!50!black, title=Prompt of Query Generation \textbf{(A - a) $\cup$ (B - b)}, fonttitle=\bfseries, fontupper=\small]
		Given an example in information retrieval tasks. We refer to the query as a negative-constraint query. The query matches formulation \textbf{(A - a) $\cup$ (B - b)}. \textbf{A} denotes Ginsberg's works, \textbf{a} denotes 'Howl', \textbf{B} denotes Poe's works and \textbf{b} denotes 'The Raven'. The positive document mentions Ginsberg's and Poe's works but does not mention 'Howl' and 'The Raven'. The negative document 1 mentions Ginsberg's works,  Poe's works' and 'Howl'. The negative document 2 mentions Ginsberg's works, Poe's works' and 'The Raven'. The negative document 3 mentions Ginsberg's works,  Poe's works', 'Howl' and 'The Raven'. Please provide a query based on the positive and negative documents provided.\\
		------\\
		\textbf{EXAMPLE}\\
		\textbf{Positive document}:\\
		Ginsberg took part in decades of non-violent political protest against everything from the Vietnam War to the War on Drugs. His poem ""September on Jessore Road"", calling attention to the plight of Bangladeshi refugees, exemplifies what the literary critic Helen Vendler described as Ginsberg's tireless persistence in protesting against ""imperial politics, and persecution of the powerless."" His collection ""The Fall of America"" shared the annual U.S. National Book Award for Poetry in 1974. In 1979 he received the National Arts Club gold medal and was inducted into the American Academy and Institute of Arts and Letters. Ginsberg was a Pulitzer. produce his own journal ""The Penn"" (later renamed ""The Stylus""), though he died before it could be produced. by gambling, and the cost of secondary education for Poe. He attended the University of Virginia but left after a year due to lack of money. Poe quarreled with Allan over the funds for his education and enlisted in the Army in 1827 under an assumed name. It was at this time that his publishing career began, albeit humbly, with the anonymous collection ""Tamerlane and Other Poems"" (1827), credited only to ""a Bostonian"". With the death of Frances Allan in 1829, Poe and Allan reached a temporary rapprochement. However, Poe later failed as an officer cadet at West Point, declaring...\\
		\textbf{Negative document 1}: \\
		Kerouac and William S. Burroughs. Ginsberg is best known for his poem ""Howl"", in which he denounced what he saw as the destructive forces of capitalism and conformity in the United States. In 1956, ""Howl"" was seized by San Francisco police and US Customs. In 1957, it attracted widespread publicity when it became the subject of an obscenity trial, as it described heterosexual and homosexual sex at a time when sodomy laws made homosexual acts a crime in every U.S. state. ""Howl"" reflected Ginsberg's own homosexuality and his relationships with a number of men, including Peter Orlovsky, his lifelong partner. by gambling, and the cost of secondary education for Poe. He attended the University of Virginia but left after a year due to lack of money. Poe quarreled with Allan over the funds for his education and enlisted in the Army in 1827 under an assumed name. It was at this time that his publishing career began, albeit humbly, with the anonymous collection ""Tamerlane and Other Poems"" (1827), credited only to ""a Bostonian"". With the death of Frances Allan in 1829, Poe and Allan reached a temporary rapprochement. However, Poe later failed as an officer cadet at West Point, declaring...\\
		\textbf{Negative document 2}: \\
		Ginsberg took part in decades of non-violent political protest against everything from the Vietnam War to the War on Drugs. His poem ""September on Jessore Road"", calling attention to the plight of Bangladeshi refugees, exemplifies what the literary critic Helen Vendler described as Ginsberg's tireless persistence in protesting against ""imperial politics, and persecution of the powerless."" His collection ""The Fall of America"" shared the annual U.S. National Book Award for Poetry in 1974. In 1979 he received the National Arts Club gold medal and was inducted into the American Academy and Institute of Arts and Letters. Ginsberg was a Pulitzer. produce his own journal ""The Penn"" (later renamed ""The Stylus""), though he died before it could be produced. a firm wish to be a poet and writer, and he ultimately parted ways with John Allan. Poe switched his focus to prose and spent the next several years working for literary journals and periodicals, becoming known for his own style of literary criticism. His work forced him to move among several cities, including Baltimore, Philadelphia, and New York City. In Richmond in 1836, he married Virginia Clemm, his 13-year-old cousin. In January 1845, Poe published his poem ""The Raven"" to instant success. His wife died of tuberculosis two years after its publication. For years, he had been planning to...\\
		\textbf{Negative document 3}: \\
		Kerouac and William S. Burroughs. Ginsberg is best known for his poem ""Howl"", in which he denounced what he saw as the destructive forces of capitalism and conformity in the United States. In 1956, ""Howl"" was seized by San Francisco police and US Customs. In 1957, it attracted widespread publicity when it became the subject of an obscenity trial, as it described heterosexual and homosexual sex at a time when sodomy laws made homosexual acts a crime in every U.S. state. ""Howl"" reflected Ginsberg's own homosexuality and his relationships with a number of men, including Peter Orlovsky, his lifelong partner. a firm wish to be a poet and writer, and he ultimately parted ways with John Allan. Poe switched his focus to prose and spent the next several years working for literary journals and periodicals, becoming known for his own style of literary criticism. His work forced him to move among several cities, including Baltimore, Philadelphia, and New York City. In Richmond in 1836, he married Virginia Clemm, his 13-year-old cousin. In January 1845, Poe published his poem ""The Raven"" to instant success. His wife died of tuberculosis two years after its publication. For years, he had been planning to...\\
		\textbf{Query}:\\
		What are the similarities between Ginsberg's works (excluding 'Howl') and Poe's works (excluding 'The Raven')?\\
		------\\
		\textbf{Positive document}: \%POSITIVE DOCUMENT\%\\
		\textbf{Negative document 1}: \%NEGATIVE DOCUMENT 1\%\\
		\textbf{Negative document 2}: \%NEGATIVE DOCUMENT 2\%\\
		\textbf{Negative document 3}: \%NEGATIVE DOCUMENT 3\%\\
		------\\
		Below query is the one you need to generate, which make significant changes to the query style.\\
		\textbf{Query}:\\
		\%QUERY\%
	\end{tcolorbox}
    \twocolumn
	\section{Data Collection and Snippet}\label{negconstraint}
	We use the introductory passages from Wikipedia dump as the corpus, as they are usually high-quality and contain most of the key information.  We request three carefully selected experienced annotators to filter passages from Wikipedia. For queries of formulation \textbf{A - a} and \textbf{(A - a) $\cup$ B}, each positive and negative document corresponding to a query is composed of one passage from the Wikipedia dump, respectively. However, due to the complexity of formulation \textbf{(A - a) $\cup$ (B - b)}, we merge two passages that belong to one topic as a positive document, and the two merged passages are highly relevant. Negative documents for formulation \textbf{(A - a) $\cup$ (B - b)} are also obtained in this way. Then we prompt GPT-4o to generate queries based on the positive and negative documents. We ensure that the queries are as style-diverse as possible. That is, we do not just perform entity replacement, but pay more attention to the diversity of query mode. For example, there are three queries expressing negative constraints for  formulation \textbf{ A - a}:
	\begin{enumerate}
		\item \emph{Investigate the role of nature in Walden, excluding Thoreau’s critique of society.}
		\item \emph{Introduce the works of Emily Dickinson, but do not mention 'Because I could not stop for Death'.}
		\item \emph{Without referencing Victor Frankenstein's use of scientific knowledge, examine the role of technology in Frankenstein.}
	\end{enumerate}
	Finally, annotators also select several irrelevant passages with queries to fill into the corpus.\\
	\indent Table~\ref{tab8} introduce snippets of NegConstraint dataset. Entities marked in red and green denote entities in negative-constraint conditions.  For formulation \textbf{A - a}, the negative document mentions "Howl". For formulation \textbf{ (A - a) $\cup$ B}, the negative document mentions "Howl". For formulation \textbf{(A - a) $\cup$ (B - b)}, the negative document 1 mentions "Howl", the negative document 2 mentions "The Raven", and the negative document 3 mentions "Howl" and "The Raven". 
	\begin{table*}[htpb]
		\small
		\centering
		\begin{tabularx}{\textwidth}{c|p{3cm}|X}
			\hline
			\multirow{8}*{\textbf{A - a}}	
			&Query&Introduce Allen Ginsberg's works, but do not mention ' \textcolor{red}{Howl}'.\\
			\cline{2-3}
			&\multirow{4}*{Postive document}&…His poem 'September on Jessore Road', calling attention to the plight of Bangladeshi refugees, exemplifies what the literary critic Helen Vendler described as Ginsberg's tireless persistence in protesting against ""imperial politics, and persecution of the powerless…\\
			\cline{2-3}
			&\multirow{3}*{Negative document}&In 1956, '\textcolor{red}{Howl}' was seized by San Francisco police and US Customs...'\textcolor{red}{Howl}' reflected Ginsberg's own homosexuality and his relationships with a number of men, including Peter Orlovsky, his lifelong partner…\\
			\hline
			\multirow{12}*{\textbf{(A - a) $\cup$ B}}	
			&\multirow{2}*{Query}&What themes are expressed in Allen Ginsberg's works other than '\textcolor{red}{Howl}' and Edgar Allan Poe’s works?\\
			\cline{2-3}
			&\multirow{6}*{Postive document}&Ginsberg took part in decades of non-violent political protest against everything from the Vietnam War to the War on Drugs. His poem ""September on Jessore Road"", calling attention to the plight of Bangladeshi refugees, exemplifies what the literary critic Helen Vendler described as Ginsberg's tireless persistence in protesting against ""imperial politics, and persecution of the powerless…\\
			\cline{2-3}
			&\multirow{4}*{Negative document}&Kerouac and William S. Burroughs. Ginsberg is best known for his poem '\textcolor{red}{Howl}', in which he denounced what he saw as the destructive forces of capitalism and conformity in the United States. In 1956, '\textcolor{red}{Howl}' was seized by San Francisco police and US Customs….\\
			\hline
			\multirow{26}*{\textbf{(A - a) $\cup$ (B - b)}}
			&\multirow{2}*{Query}	&What themes do Allen Ginsberg's works other than '\textcolor{red}{Howl}' and Edgar Allan Poe's works other than '\textcolor{green}{The Raven}' express?\\
			\cline{2-3}
			&\multirow{8}*{Postive document}&Ginsberg took part in decades of non-violent political protest against everything from the Vietnam War to the War on Drugs..., His collection 'The Fall of America' shared the annual U.S. National Book Award for Poetry in 1974. In 1979 he received the National Arts Club gold medal and was inducted into the American Academy and Institute of Arts and Letters...,credited only to 'a Bostonian'. With the death of Frances Allan in 1829, Poe and Allan reached a temporary rapprochement. However, Poe later failed as an officer cadet at West Point, declaring...\\
			\cline{2-3}
			&\multirow{6}*{Negative document 1}&Kerouac and William S. Burroughs. Ginsberg is best known for his poem '\textcolor{red}{Howl}', in which he denounced what he saw as the destructive forces of capitalism and conformity in the United States...'\textcolor{red}{Howl}' reflected Ginsberg's own homosexuality and his relationships with a number of men, including Peter Orlovsky, his lifelong partner. by gambling, and the cost of secondary education for Poe...\\
			\cline{2-3}
			&\multirow{3}*{Negative document 2}&…In January 1845, Poe published his poem '\textcolor{green}{The Raven}' to instant success. His wife died of tuberculosis two years after its publication. For years, he had been planning to...\\
			\cline{2-3}
			&\multirow{7}*{Negative document 3}&In 1956, '\textcolor{red}{Howl}' was seized by San Francisco police and US Customs... '\textcolor{red}{Howl}' reflected Ginsberg's own homosexuality and his relationships with a number of men, including Peter Orlovsky, his lifelong partner. a firm wish to be a poet and writer, and he ultimately parted ways with John Allan...In January 1845, Poe published his poem '\textcolor{green}{The Raven}' to instant success. His wife died of tuberculosis two years after its publication. For years, he had been planning to\\
			\hline
		\end{tabularx}
		\caption{Snippets of NegConstraint dataset. Entities marked in red and green denote negative-constraint conditions.}\label{tab8}
	\end{table*}
\end{document}